\documentclass[twocolumn,preprintnumbers,amsmath,prc,amssymb]{revtex4}

\usepackage{graphicx}
\usepackage{dcolumn}
\usepackage{bm}
\usepackage{amsmath}
\usepackage{lipsum}

\begin{document}

\title{The Flow Constraint Influence on the Properties of Nuclear Matter Critical Endpoint}

\author{A. I. Ivanytskyi$^1$, K. A. Bugaev$^1$, V. V. Sagun$^{1,2}$, L.V. Bravina$^{3}$ 
            and E. E. Zabrodin$^{3,4,5}$}

\affiliation{$^1$Bogolyubov Institute for Theoretical Physics of the National Academy of Sciences of Ukraine, Metrologichna str. 14$^b$, 03680 Kiev, Ukraine}
\affiliation{
$^2$Centro Multidisciplinar de Astrofisica, Instituto Superior Tecnico, Universidade de Lisboa, Av. Rovisco Pais 1, 1049-001 Lisboa, Portugal}
\affiliation{$^3$Department of Physics, University of Oslo, Sem S\ae lands vei 24, 0371 Oslo, Norway}
\affiliation{$^4$National Research Nuclear University (MEPhI), Kashira Highway 31, 115409 Moscow, Russia}
\affiliation{$^5$Skobeltsyn Institute of Nuclear Physics, Lomonosov Moscow State University, 
                           Leninskie gory, GSP-1, 119991 Moscow, Russia}



\begin{abstract}
We propose a novel family of  equations of state  for symmetric nuclear matter based on the induced surface tension  concept for the hard-core repulsion. It is shown that having only four adjustable parameters the suggested  equations of state can, simultaneously, 
reproduce not only the main properties of the nuclear matter ground state, but  the proton flow constraint up its  maximal particle number densities.  Varying the model  parameters we carefully examine the range of  values of incompressibility constant of normal nuclear matter and its critical temperature which are consistent with the proton flow constraint. This analysis allows us to show 
that the 
physically most justified value of  nuclear matter  critical temperature  is 15.5-18 MeV,  the incompressibility constant is 270-315 MeV and the hard-core radius of nucleons  is  less than 0.4 fm. \\

\noindent
{\small Keywords: Induced surface tension,  symmetric nuclear matter, proton flow constraint}

\end{abstract}

\maketitle


\section{Introduction}
\label{Intro}


The determination of   basic  characteristics of symmetric  nuclear  matter   and possible interrelations 
between them is of fundamental  importance   \cite{Horst86,Karnauch06,Lattimer12,StellarSMM13,Dutra14,Maslov17} not only for  nuclear spectroscopy and for nuclear physics of intermediate energies, but also for  nuclear astrophysics 
in view of  possible phase transformations in compact astrophysical  objects (neutron stars,  hypothetical hybrid and  quark stars). 
From the practical point of view such characteristics of infinite   nuclear  matter  as the normal density $n_0$ at zero pressure and zero temperature,  its binding energy per nucleon $W_0$  and its incompressibility factor $K_0$ are  of great importance for various 
phenomenological models because just these characteristics  are used to fix the model parameters. Furthermore,  such parameters 
of the nuclear liquid-gas transition  phase diagram as the critical temperature $T_c$,  the critical particle number  density $n_c$ and critical  pressure $p_c$ at the endpoint  and the values of  critical exponents   are  important  not only for the theory of critical phenomena, but they are also important  for  a  verification of  the novel theoretical  approaches to study  the phase transitions in   
finite systems with strong interaction \cite{Dutra14, KAB07,BISO:13,Sagun2014a,Bugaev07rev,Gulminelli15}.

Although some of these parameters, namely $n_0$ and  $W_0$ are known well, the model independent experimental determination of all other aforementioned  characteristics is extremely  difficult, since
these parameters correspond to an infinite nuclear matter, while in the experiments one can  study only the nuclei of finite size.  Therefore, any  relations or conditions which connect these characteristics are very important both for nuclear theory and for experiment.  Recently, a comprehensive analysis  of    relation between the critical temperature of hot nuclear matter  and  incompressibility factor  of  its ground state, i.e. at the particle number density  $n_0$ and vanishing temperature,  was performed 
in  Refs. \cite{Delfino16, Menezes17} for relativistic mean-field  (RMF) models.  
One of  the important constraints imposed on   the RMF  models discussed in  \cite{Dutra14,Delfino16}  is  the so-called proton flow constraint \cite{Danielewicz}.   This constraint \cite{Danielewicz} 
requires that  at vanishing temperature  and  high baryonic charge densities the realistic equations of state (EoS) are soft, i.e. it 
sets rather strong restrictions on the particle number density dependence of  pressure from two to about five values of  normal nuclear density. As a result, even having about 10 or more adjustable parameters only 104  RMF models  out of  263    analyzed  in  \cite{Dutra14}  are able to obey this constraint. It is clear that so many model parameters  do not allow to perform a systematic study of the  flow constraint influence on the characteristics  of  symmetric  nuclear  matter critical endpoint (CEP)  for  the RMF models.

At the same time,    two novel approaches   to account for the hard-core repulsion in relativistic quantum gases were suggested recently \cite{Vovch17,Bugaev2017}. 
 Their  advantage is that 
the novel EoS  allow one to go beyond the usual Van der Waals approximation \cite{Vovch17,Bugaev2017}.
However, the EoS  developed in Ref. \cite{Vovch17} employs  the parameterizations of  attractive interaction which 
are typical for classical gases  and, as a result,  even the  minimal value obtained  for  the  incompressibility factor $K_0$ is somewhat above  its  experimental range of nuclear matter \cite{Dutra14,Delfino16, Menezes17, ExpK0}, while the values of nucleon hard-core radius are too large.
Furthermore,  in  \cite{Bugaev2017} it is shown that for same  parameterization of the mean-field attractive potential and   temperatures  below  1 MeV  the EoS which belong to the class suggested in \cite{Bugaev2017} are essentially softer than their 
analogs developed in   \cite{Vovch17}.  Therefore, in order to  study the influence of the  proton  flow constraint  it is natural to 
use a softer EoS from the class  suggested in  \cite{Bugaev2017}.
For this purpose here  we formulate a family  of   4-parametric EoS with the phenomenological attraction similar to   Ref.  \cite{Gorenstein93}  which are normalized  to the properties of nuclear matter ground state and obey the proton flow constraint.  Using this EoS family, 
we perform a systematic investigation of  restrictions on the critical temperature $T_c$ and incompressibility factor 
$K_0$  generated by the flow constraint  \cite{Danielewicz}.  This study allows us to show that the critical compressibility  
factor $Z_c$ of nuclear matter can be essentially lower than the typical values $0.28-0.31$ obtained by the RMF models  \cite{Menezes17} and, hence, it can be similar to the $Z_c$ values  of  ordinary non-organic liquids. 
Based on these results, we believe that  the present approach enables us to make a bridge between the  nuclear matter EoS
and the ones for  ordinary liquids.  

The work  is organized as follows. The main ingredients of  a novel  EoS are given  in Section II. 
Section III  is devoted to a systematic analysis of the proton flow constraint influence  on the nuclear matter  EoS   and 
its CEP properties. Our conclusions are given in Section \ref{Concl}.


\section{Equation of state}
\label{EoS}

Since we develop a phenomenological model of nuclear matter, we are not bound by the Lagrangian choice and, hence, we consider only the nucleons 
assuming that effect of the $\Delta$ and  heavier  baryonic resonances which can appear at high densities  is absorbed  in the mean-fields. 
The model pressure $p$ is a solution of the system  ($R$ is the hard-core radius on nucleons)
\begin{eqnarray}
\label{I}
p&=&p_{id}(T,\nu_p) - p_{int}\bigl(n_{id}(T,\nu_p)\bigl)\,,\\
\label{II}
\Sigma&=&R\, p _{id}(T,\nu_\Sigma)\,,
\end{eqnarray}
where $p_{id}(T,\mu)$  is the grand canonical  pressure of noninteracting point-like fermions
\begin{eqnarray}
\label{III}
p_{id}(T,\nu)=Tg\int\frac{d^3p}{(2\pi)^3}\ln\left[{\textstyle 1+\exp\left(\frac{\nu-\sqrt{p^2+m^2}}{T}\right)} \right]\,,
\end{eqnarray}
and  the particle number  density is defined as 
\begin{eqnarray}
\label{IV}
n_{id}(T,\nu)= \frac{\partial p_{id}}{\partial\,  \nu}=g\int\frac{d^3p}{(2\pi)^3}\frac{1}{\exp\left(\frac{\sqrt{p^2+m^2}-\nu}{T}\right)+1}\,.
\end{eqnarray}
Here the system temperature  is $T$,  $m=940$ MeV is the nucleon mass and  the nucleon degeneracy factor is $g=4$.

The term $-p_{int}$ in Eq. (\ref{I}) represents the mean-field contribution to the pressure caused by an attraction between the nucleons. 
The quantity $\Sigma$ in Eq. (\ref{II}) is the surface tension induced by the hard-core repulsion between the nucleons and, hence,  in Ref. \cite{Sagun2014a} it was called as the induced surface tension (IST) in order to distinguish it from the eigensurface tension of ordinary nuclei.  Its meaning as the surface tension coefficient can be easily seen from the effective chemical potentials
which  are  defined through the baryonic  chemical potential $\mu$ as
\begin{eqnarray}
\label{V}
\nu_p&=&\mu-pV_0-\Sigma S_0+U\bigl(n_{id}(T,\nu_p)\bigl)\,,\\
\label{VI}
\nu_\Sigma&=&\mu-pV_0 -\alpha\Sigma S_0 +U_0\, ,
\end{eqnarray}
where $V_0=\frac{4\pi}{3}R^3$ and $S_0=4\pi R^2$ are, respectively,  the eigenvolume and the eigensurface  of a  particle with the hard-core  radius $R$,
and the attractive mean-field potentials are  $U\bigl(n_{id}(T,\nu_p)\bigl)$ and $U_0 = const$.
The system (\ref{I})-(\ref{VI}) is a concrete realization of the quantum model suggested in  \cite{Bugaev2017}, where the self-consistency condition 
\begin{equation}
\label{VII}
p_{int}(n)=n\, U(n)- \int_0^n  dn' \, U(n') \,,
\end{equation}
was thoroughly  discussed for the EoS of the same class  as the one defined by Eq.  (\ref{I})-(\ref{VI}).  Eq. (\ref{VII})  relates the interaction pressure $p_{int}\bigl(n_{id}(T,\nu_p)\bigl)$ and the corresponding mean-field potential $U\bigl(n_{id}(T,\nu_p)\bigl)$ and it guarantees the fulfillment of all thermodynamic identities \cite{Bugaev2017}.  

It is worth to note  that substituting  the  constant potential $U_0\bigl(n_{id} (T,\nu_\Sigma) \bigr)  = const$ into the consistency condition  (\ref{VII}),  one automatically obtains that  the corresponding  mean-field pressure should be zero, i.e. $\tilde p_{int}\bigl(n_{id}(T,\nu_\Sigma)\bigl)=0$.  
We would like to note that different density dependence  of the attractive mean-field potentials $U\bigl(n_{id})$ and $U_0$
reflects   the different origins  of their forces, namely $U\bigl(n_{id})$ is generated by the bulk part of interaction, while $U_0$ is
attributed to the surface part. The meaning of $U_0$ potential can be understood  after  the non-relativistic  expansion of the 
particle energy $\sqrt{m^2 + k^2} \simeq m +  \frac{k^2}{2 m}$ in the momentum  distribution function in Eq. (\ref{IV}): 
$U_0$ decreases the nucleon mass to the value $m - U_0$. 

Finding the partial $\mu$ derivatives 
of  Eqs.  (\ref{I}) and  (\ref{II}), one can get
the particle number density from the usual thermodynamic identity 
\begin{equation}
\label{VIII}
n= \frac{\partial p}{\partial \mu} = \frac{n_{id}(T,\nu_p)}{1+V_0~n_{id}(T,\nu_p) + \frac{3\, V_0\,n_{id}(T,\nu_\Sigma) }{1+3(\alpha-1)V_0\, n_{id}(T,\nu_\Sigma)}}\, . 
\end{equation}

In principle, Eq.  (\ref{II}) for  the IST coefficient could contain the interaction pressure $\tilde p_{int}(n_{id} (T, \nu_\Sigma)$ \cite{Bugaev2017}. However, since the pioneering work  \cite{Rischke91}, in which the Van der Waals-like hard-core repulsion, i.e. 
the term $-p V_0$ in Eq. (\ref{V}), 
was introduced into the RMF model of  nuclear matter, it is well known that such a repulsion is very weak at the vicinity of the nuclear matter ground state because in this region $p \simeq 0$ and, hence, an additional repulsion is absolutely  necessary.   In Ref.  \cite{Rischke91} the additional repulsion was provided by a vector meson field interacting with nucleons, while here such a repulsion is exclusively  provided by the  IST  coefficient  $\Sigma$. Hence, its 
interaction pressure $\tilde p_{int}(n_{id} (T, \nu_\Sigma) )$  could not contain any attraction in contrast to the term $p_{int}(n_{id}(T, \nu_p))$ in  Eq. (\ref{I}).  As it was shown in  \cite{Bugaev16,Sagun17,Bugaev17b} exactly the form of   Eq. (\ref{II}), i.e. with 
$\tilde p_{int}(n_{id} (T, \nu_\Sigma) ) \equiv 0 $, allows one to correctly account for the hard-core repulsion in  case of the Boltzmann statistics 
up to the packing fractions $\eta \equiv V_0 n \simeq 0.2 $ (here $n$ is the particle number density), if the parameter $\alpha$ is chosen as $\alpha= 1.245$. An additional reason   for  such a simple parameterization of  Eq. (\ref{II}) is to keep the number of parameters as small as possible. Due to the same reason  for the present model we fix   $\alpha= 1.245$.

The role of  the parameter $\alpha =1.245$ can be seen from the expression for the particle number density (\ref{VIII}). 
Indeed,
from Eq.  (\ref{VIII})  one can see that at low pressures, when the excluded volume effects  are weak and the system is close to the non-relativistic ideal gas, i.e. for $\nu_p \ll m$, $\nu_\Sigma \ll m$  and the temperatures $|\nu_p - \nu_\Sigma| \ll T \ll m$, then the  densities  $n_{id}(T,\nu_p)$ and $n_{id}(T,\nu_\Sigma)$ are simply equal to each other, i.e. 
$n_{id}(T,\nu_p) \simeq  n_{id}(T,\nu_\Sigma)$,  and, hence, the particle number density  $n \simeq \frac{n_{id}(T,\nu_p)}{1+4\,V_0\, n_{id}(T,\nu_p) }$ acquires the typical  one component excluded volume (EV)  form \cite{Zeeb08}.  The last equality was obtained from Eq.  (\ref{VIII})   under an evident approximation that at low pressures and densities  the term $V_0\, n_{id}(T,\nu_\Sigma) \ll 1$  is small and, hence,  it  can  be neglected. Thus, at low pressures the system (\ref{I})-(\ref{VI}) recovers the usual EV results by construction. 

At higher pressures the situation is defined by the value of parameter $\alpha$.
If $\alpha < 1$, then at some value of  $ n_{id}(T,\nu_\Sigma) = \frac{1}{3(1-\alpha) V_0} > 0$ the particle number density vanishes and further increase of pressure  makes it negative. Hence, we conclude that the case    $\alpha < 1$ is unphysical. 
If $\alpha > 1$,  then at high pressures both densities $n_{id}(T,\nu_p)$ and $n_{id}(T,\nu_\Sigma)$ diverge and, hence, the particle number density  is  $n \rightarrow 1/V_0$, i.e. it  is equal to the inverse value of  nucleon eigenvolume. This feature of the present EoS is caused by an accurate parameterization  of the hard-core repulsion effects for $\alpha > 1$. Hence,  fixing $\alpha =1.245$ we keep 
the connection to the  results  obtained for the Boltzmann   statistics at high temperatures \cite{Bugaev16,Sagun17,Bugaev17b}.

If, however, $\alpha =1$, then at high pressures  the behavior of  particle number density  $n = \frac{n_{id}(T,\nu_p)}{1+V_0~n_{id}(T,\nu_p) + 3\, V_0\,n_{id}(T,\nu_\Sigma) }$  strongly  depends  on the details of the model  interaction. Thus, for $\mu \rightarrow \infty$ one finds  that $\nu_p - \nu_\Sigma = U (n_{id}(T,\nu_p)) - U_0$. If the function  $U (n_{id}(T,\nu_p)) $ corresponds to an attraction and it is a growing function of its argument $n_{id}(T,\nu_p)$, then in this limit one finds $n_{id}(T,\nu_p) \gg n_{id}(T,\nu_\Sigma)$ and, hence,  one finds  $n \rightarrow 1/V_0$.  Apparently, 
the speed of approaching the limiting value depends on the strength of mean-field potentials $U (n)$ and $U_0$. Now it is  clear that for the case of repulsion, i.e.  for $U (n_{id}(T,\nu_p))  < 0$,  the particle number density $n \rightarrow \frac{n_{id}(T,\nu_p)}{3\, V_0\,n_{id}(T,\nu_\Sigma) } \ll 1/V_0$, i.e. it  can be even lower,  than 
for the classical EV case. This is, actually, one of the reasons of why  the repulsion of the present model is exclusively 
described by the hard-core repulsion which allows one to avoid such problems for $\alpha > 1$.


\section{Nuclear matter properties}
\label{NMP}

In this work we use the power parameterization of the mean-field potential motivated by Ref. \cite{Gorenstein93}. i.e.
\begin{eqnarray}
\label{IX}
U(n)=C_d^2 n^\kappa\, \quad \Rightarrow  \quad  p_{int}(n)=\frac{\kappa}{\kappa+1}C_d^2 n^{\kappa+1}\, ,
\end{eqnarray}
where  the mean-field contribution to the pressure $p_{int}(n)$ is obtained from the consistency condition (\ref{VII}).
Note that this is one of the simplest choices of  the mean-field potential which includes two parameters only, i.e. $C_d^2$ and $\kappa$.  Since the parameter $\alpha$ is fixed the other two parameters of the IST model are the hard-core radius $R$ and the 
constant potential $U_0$.     Also it is important that  in a general way one can show that in contrast to other phenomenological EoS the present one obeys  the Third Law of thermodynamics \cite{Bugaev2017}. 
\vspace*{-0.4mm}
\begin{figure}[th]
\centerline{\includegraphics[width=0.95\columnwidth]{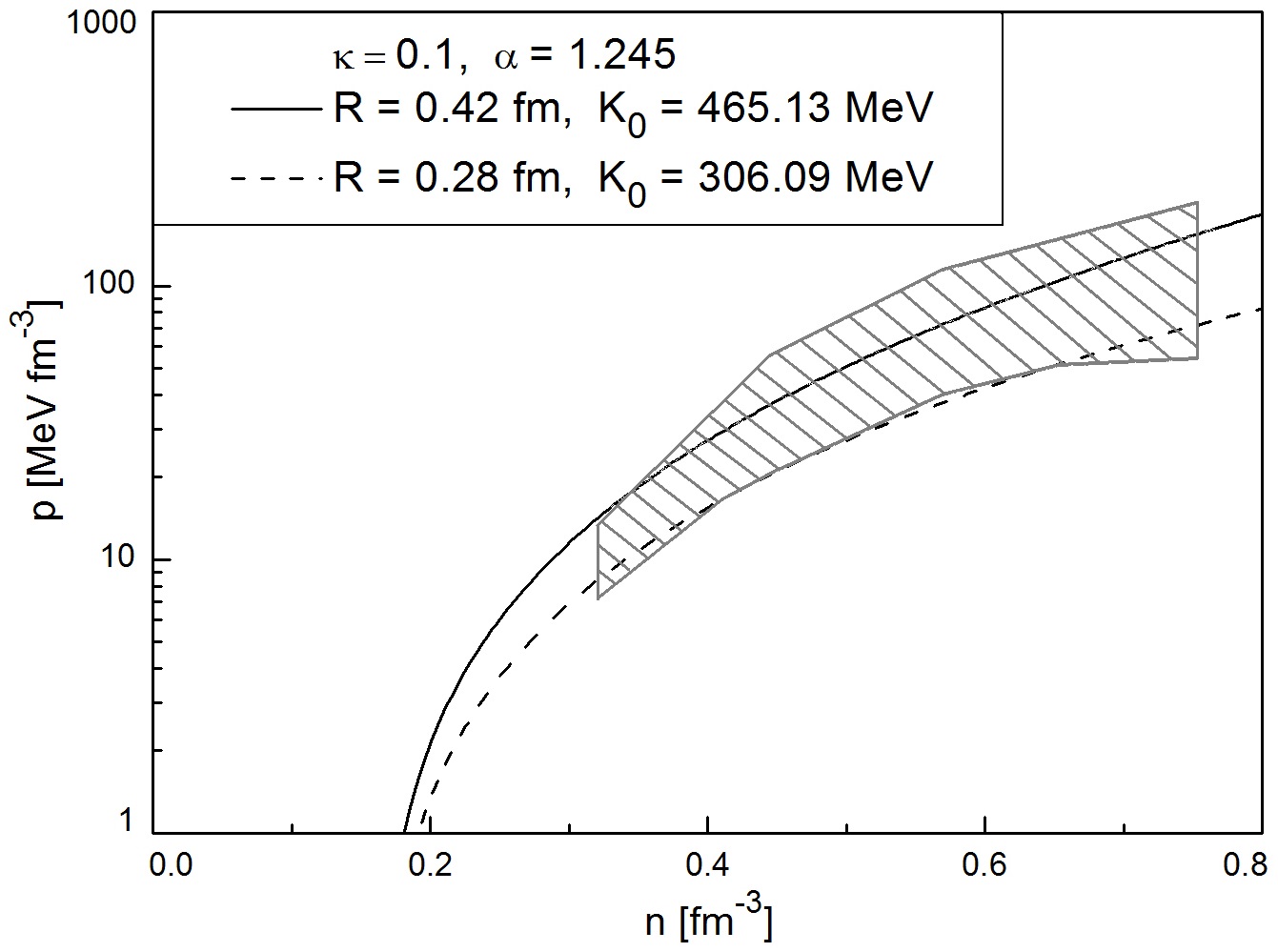}}
\centerline{\includegraphics[width=0.95\columnwidth]{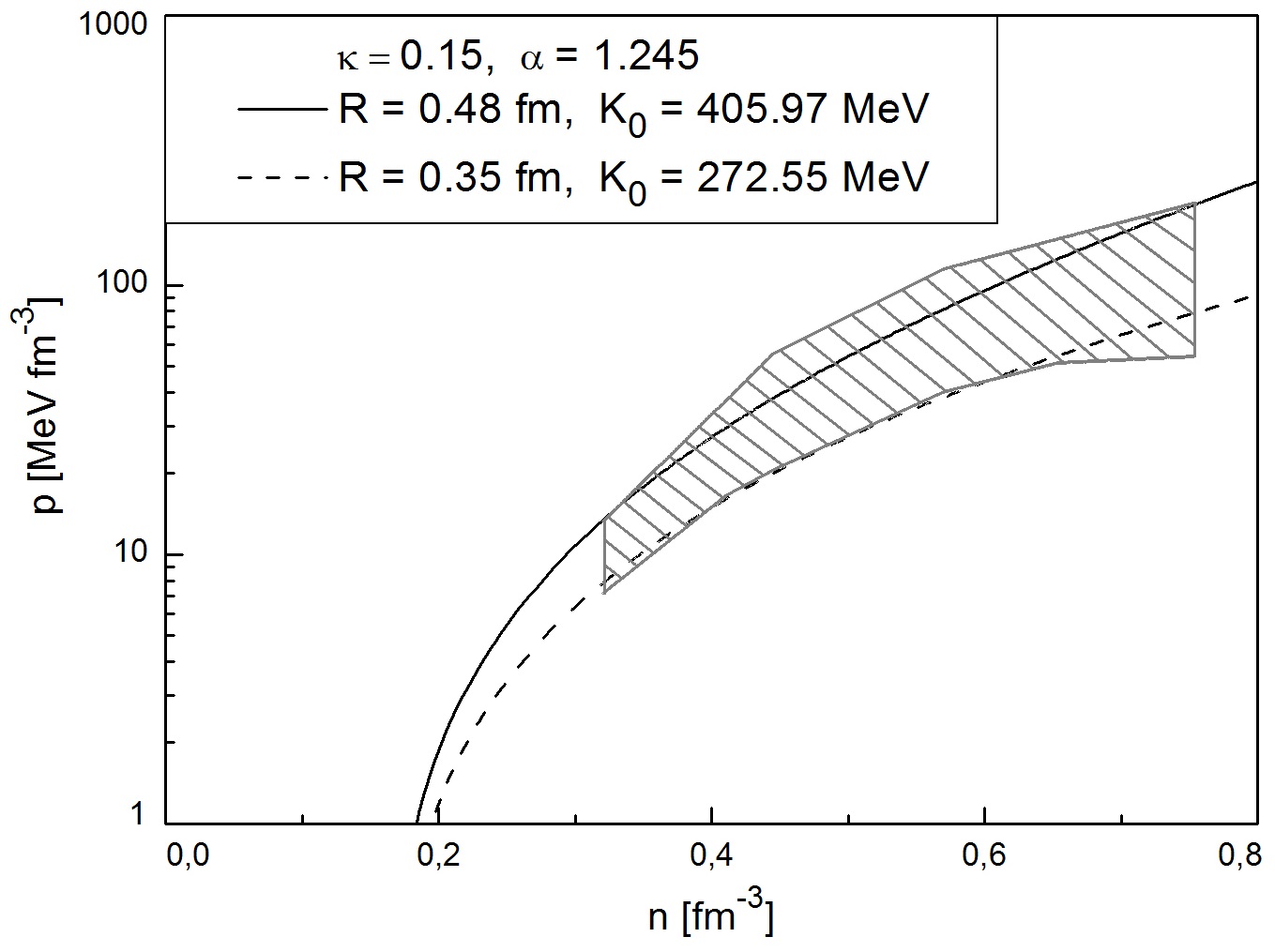}}
\caption{Density dependence of the system pressure is shown for  several set of parameters
which are specified in the legend of each panel. See Table I for more details. 
The dashed area corresponds to the  proton flow constraint of Ref.  \cite{Danielewicz}
}
\label{fig1}
\end{figure}
\begin{figure}[th]
\centerline{\includegraphics[width=0.95\columnwidth]{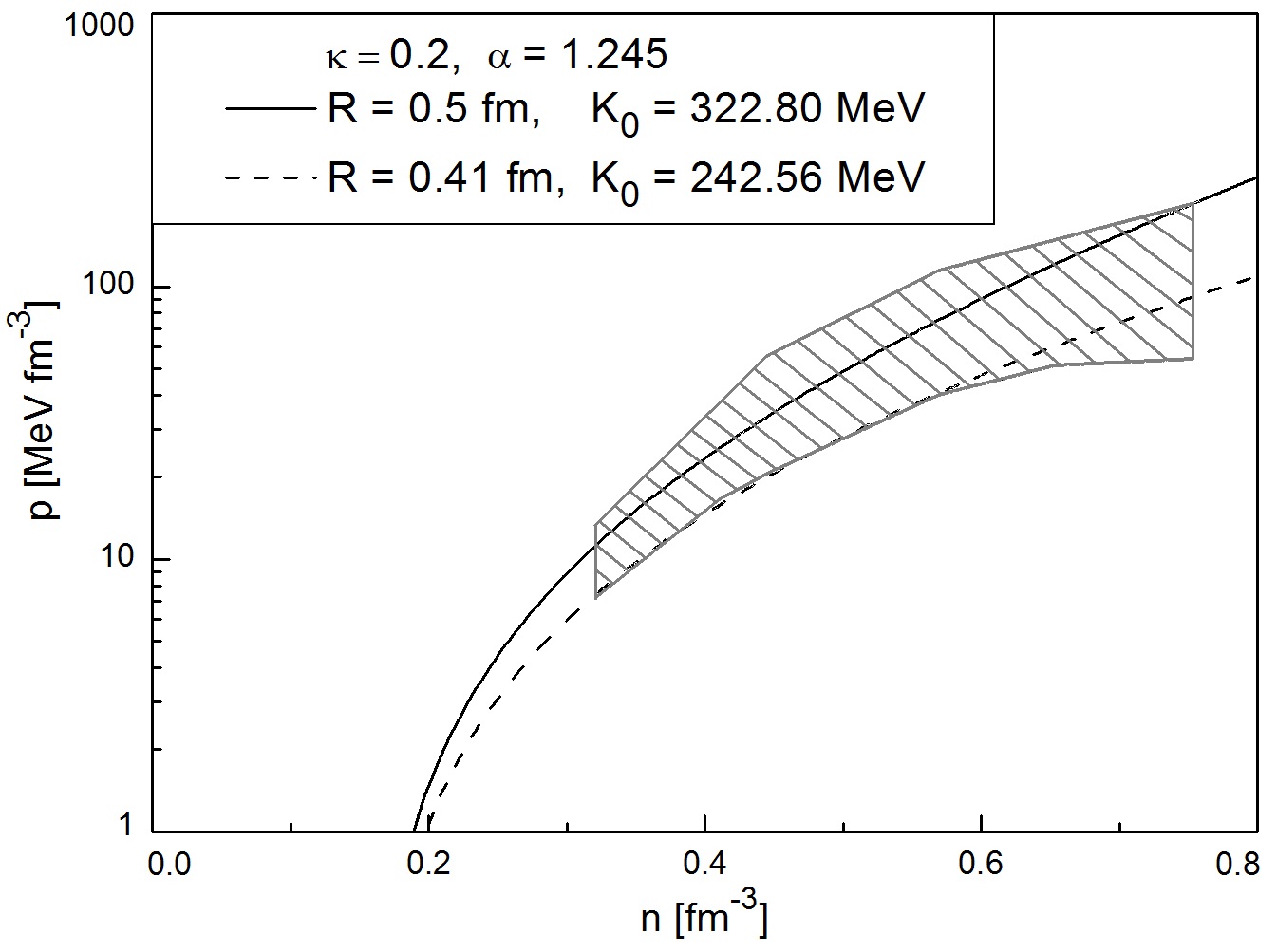}}
\centerline{\includegraphics[width=0.95\columnwidth]{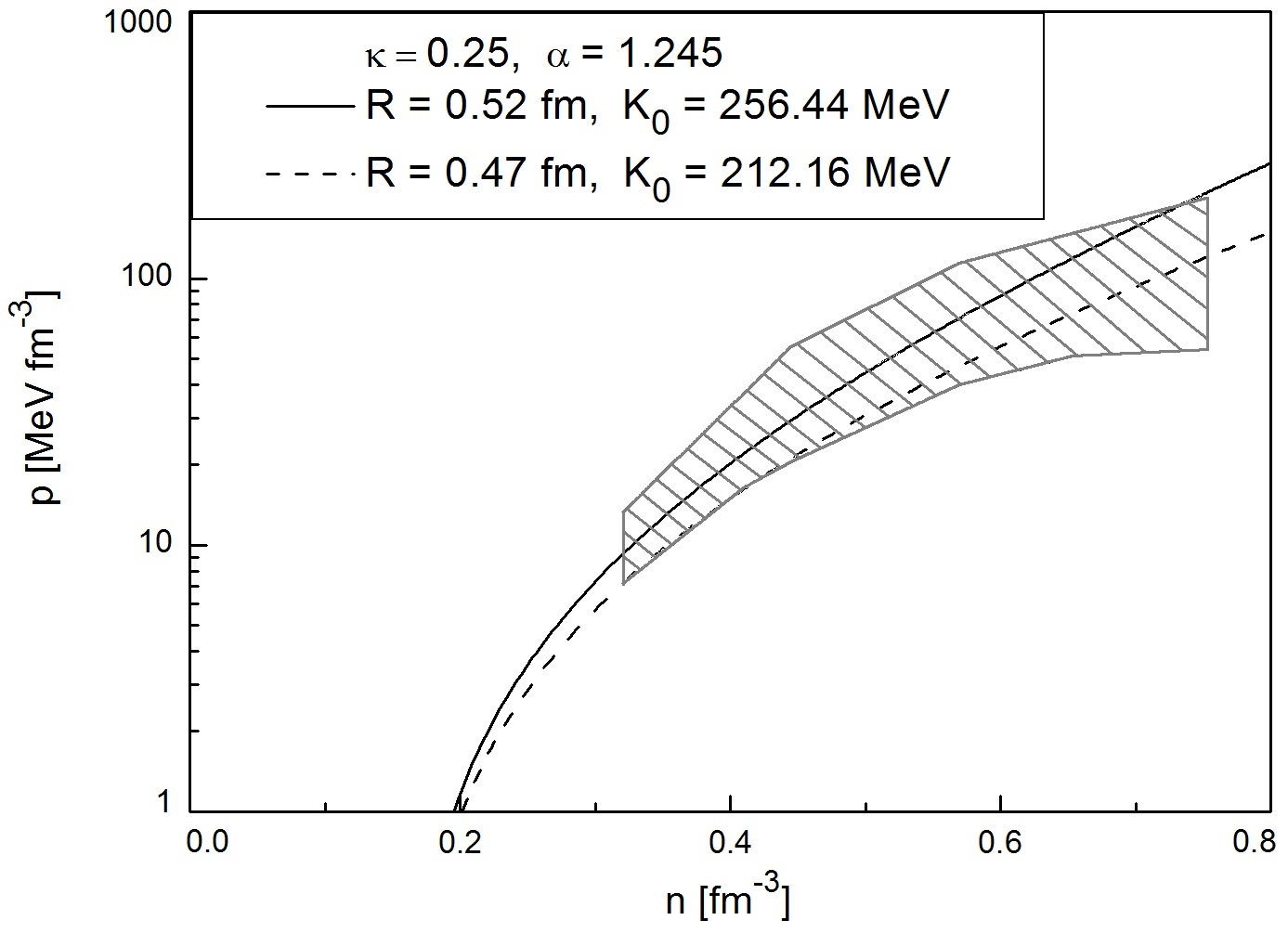}}
\centerline{\hspace*{2.8mm}\includegraphics[width=0.91\columnwidth]{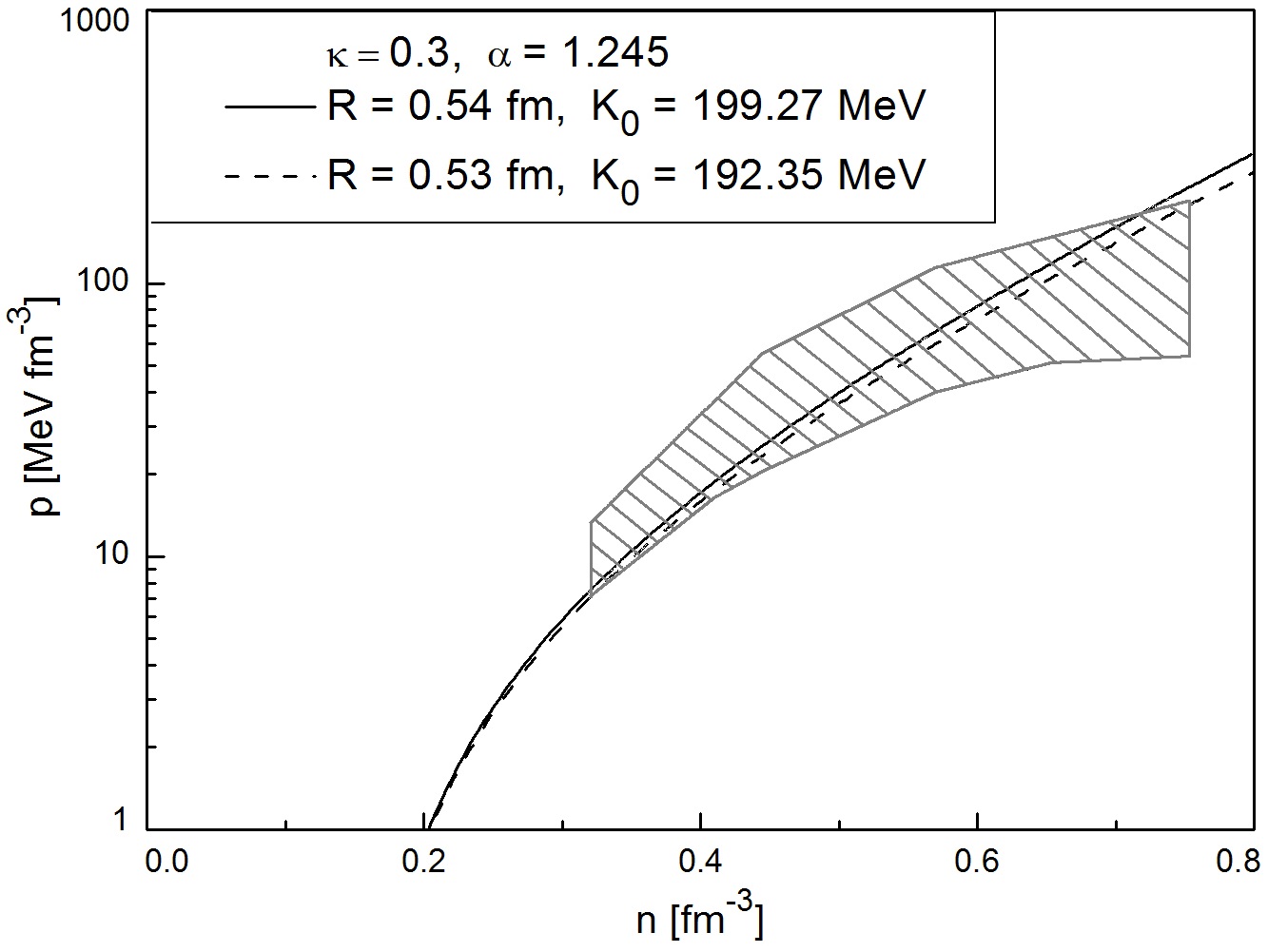}}
\caption{Same as in Fig. \ref{fig1}, but for $\kappa =0.2, 0.25$ and $0.3$.}
\label{fig2}
\end{figure}

The IST EoS with four adjustable parameters  allows one to simultaneously reproduce the ground state properties 
of symmetric nuclear matter, i.e. it has  a vanishing pressure $p=0$ at zero temperature and the normal nuclear particle number density $n_0 = 0.16$ fm$^{-3}$ and the value of its binding energy per nucleon $W_0 = \frac{\epsilon}{n} - m = -16$ MeV  (here $\epsilon$ denotes the energy density) and, hence, the corresponding chemical potential is $\mu = 923$ MeV.
The present EoS  with the attraction term (\ref{IX}) was normalized to these properties of nuclear matter ground state  for  several
values of parameter $\kappa=0.1,~0.15,~0.2,~0.25$ and $0.3$  and, simultaneously, it was fitted to obey the proton flow constraint.  For a fixed value of parameter $\kappa$ the two curves in the $n-p$ plane  were found in such a way that the upper curve 
is located not above  the upper branch of the flow constraint, while the lower one  is located not below the lower branch of this constraint. The details are clear from Figs. \ref{fig1} and  \ref{fig2}.
This is highly nontrivial results for an EoS with only four adjustable parameters, since to parameterize the proton flow constraint  alone one needs at least 8 independent points! One can readily check that all parameterizations of the IST EoS shown in 
Figs. \ref{fig1} and  \ref{fig2} also obey the kaon production constraint  obtained in Ref. \cite{KaonConstr} for the 
symmetric nuclear matter pressure in the following range $ 1.2 \, n_0 < n < 2.2 \, n_0$ of  the particle number density $n$. 

The larger values of parameter $\kappa$ were not considered, since the good description
of the proton flow constraint cannot be achieved for $\kappa \ge 0.33$.  The reason is apparent from the lower panel of Fig. \ref{fig2}. 
The values of parameter $\kappa$ below 0.1 were not considered as well because they correspond to very  large values of the incompressibility constant $K_0 \equiv 9\frac{\partial p}{\partial n}\bigl|_{T=0,~n=n_0}$.   As one can see from Table I for  $\kappa=0.1$ the minimal value of the incompressibility constant $K_0$ is about 306 MeV, while for  $\kappa < 0.1$ it gets even larger.  


\begin{figure}[th]
\vspace*{0.0mm}
\centerline{\includegraphics[width=0.99\columnwidth]{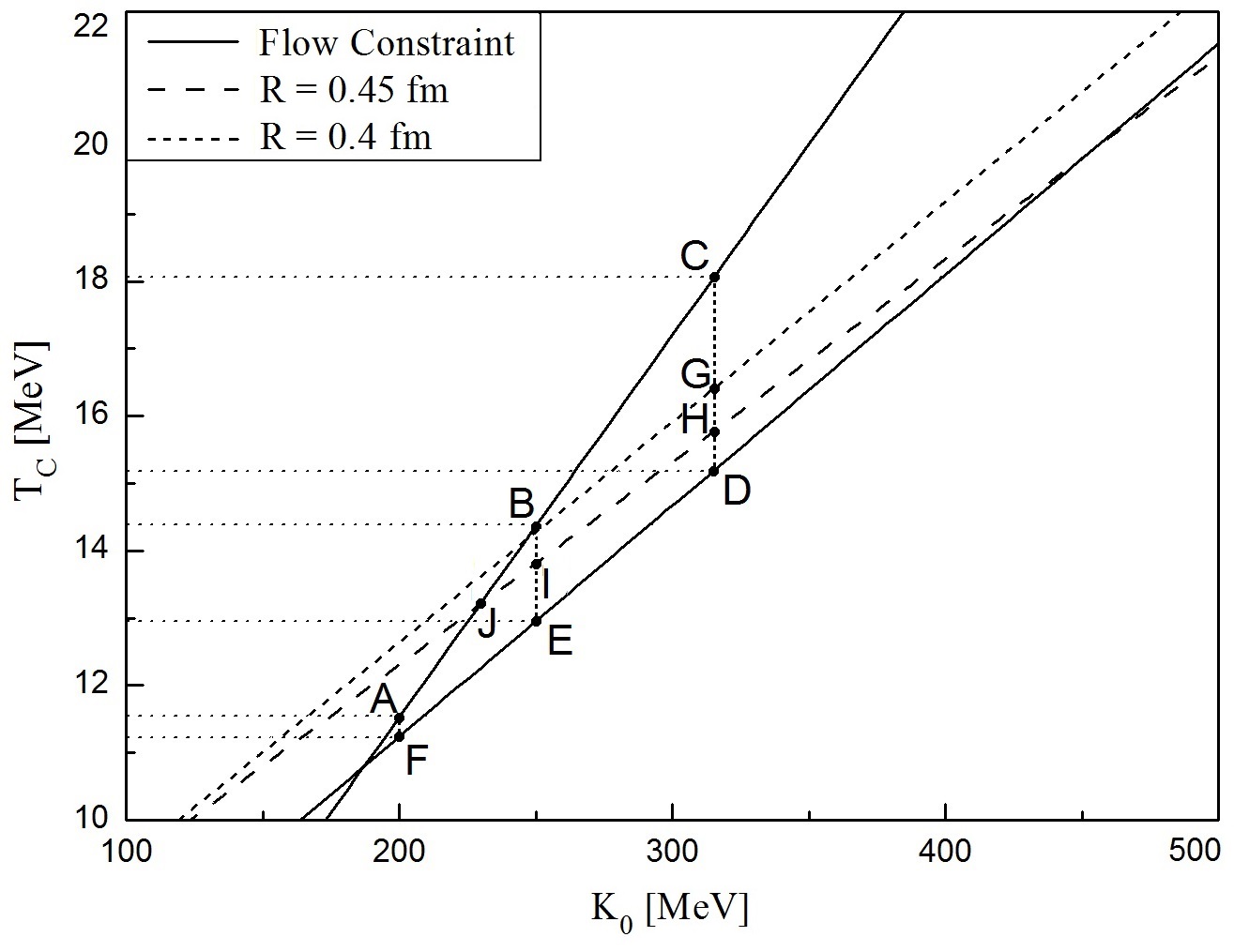}}
\caption{Values of incompressibility constant $K_0$ and critical temperature $T_C$ which obey the proton  flow constraint
are located between the lines ABC and FED. 
The lines ABC and FED are, respectively,  generated by the lower and upper branches of the proton  flow constraint.
The vertical lines AF, BE and CD correspond to $K_0$ values 200 MeV, 250 MeV and 315 MeV, respectively. }
\label{fig3}
\end{figure}

Of course, we  employed the other parameterizations of the attractive mean-field potential $U(n)$, namely the Van der Waals one   
$U(n) = 2 a n$, the constant one  $U(n) = c$ and the Clausius one $U(n)= \frac{a}{c}\left(1-\frac{1}{(1+c\, n)^2}\right)$ with the constant values of  parameters $a$ and $c$, but  
none of them gave as good results, as we found for the parameterization (\ref{IX}) with  $\alpha=1.245$.    Therefore, we believe that the IST EoS with the attraction (\ref{IX}) catches the correct physics from the normal nuclear density up to the maximal particle number density $ n_{max} \simeq 0.75$ fm$^{-3}$ of the proton flow  constraint.

The obtained  results are summarized  in Fig. \ref{fig3} and in Table I.  In Fig.  \ref{fig3} we divided the range of  $K_0$ values
into two regions, namely the lower one  $K_0= [200, 250]$ MeV and the upper one $K_0= [250, 315]$ MeV. The lower region
of  $K_0$ values corresponds to the traditional  experimental estimates (see a discussion in \cite{Dutra14}), while the upper one corresponds to the more recent estimates given in \cite{ExpK0}.  The proton flow constraint defines the allowed  region 
of $K_0$  and critical temperature $T_c$ values which are located between the lines ABC and FED in Fig.  \ref{fig3}.
From  Fig.  \ref{fig3} one can see that the lower region of  $K_0$ values determines  the rectangle ABEF for the corresponding
$T_c$ values, while the upper one determines the rectangle BCDE.  The obtained range of values is very similar to the results of  RMF models and  the non-relativistic  mean-field ones discussed in  \cite{Delfino16}. 

However,  the IST EoS allows one to obtain an essentially narrower  range of $K_0$ and $T_c$ values. Indeed, if one requires that this EoS should be applicable at the maximal value of particle number density $ n_{max} \simeq 0.75$ fm$^{-3}$ of the proton flow  constraint, then such a condition  acquires   the form
\begin{eqnarray}\label{EqX}
\frac{4}{3} \pi R^3 n_{max}~ \le~ \eta_{max} \,,
\end{eqnarray}
where the range of the model applicability is given by the maximal packing fraction $\eta_{max}$ of the model. 
Assuming that the maximal packing fraction of the present model is $\eta_{max} =0.2 $, i.e. it is  similar to the Boltzmann version of the 
IST EoS \cite{Bugaev16,Sagun17,,Bugaev17b}, one finds the following inequality on the nucleon hard-core radius $R\le 0.4$ fm. This border line is shown in  Fig.  \ref{fig3} by the short dashed line BG.  It is necessary to stress that  the value 0.4 fm is only 10\% larger than the hard-core radius of  baryons recently determined within  the IST formulation of the hadron resonance gas  model  from fitting 
the hadronic multiplicities measured in central nuclear collisions at  the AGS, SPS, RHIC and LHC energies \cite{Bugaev16,Sagun17,Bugaev17b}. 

If, however,  the present model has a wider range of applicability, i.e. $\eta_{max} =0.3$, then the inequality for  the nucleon hard-core radius becomes  $R\le 0.45$ fm. It is shown in Fig.  \ref{fig3} by the long dashed line JH. Since there is no reason to expect that
the quantum version of the IST EoS is applicable at the packing fractions exceeding the value $\eta_{max} = 0.3$ we consider it as an upper  limit of the model applicability.  Alternatively, this means that the value  $0.45$ fm is an  upper limit for the hard-core radius of nucleons. 
  
  The weak radius constraint $R\le 0.45$ fm immediately reduces the range of $K_0$ and $T_c$ values to the triangle JCH in Fig.  \ref{fig3}. 
The strong radius constraint $R \le 0.4$ fm defines even smaller triangle BCG of the allowed $K_0$ and $T_c$ values in 
Fig.  \ref{fig3}. 
Note that for the constraint $R\le 0.45$ fm the lower range of $K_0$   values gets narrower, i.e.  $K_0 \in [230; 250]$ MeV and, hence,  $T_c \in [13.2; 14.3]$ MeV, while for the inequality $R\le 0.4$ fm there are not  allowed values of  $K_0$  from the lower 
range of values as one can see from  Fig.  \ref{fig3}.  In other words, the constraint $R\le 0.4$ fm rules out the values of the incompressibility $K_0 < 250$ MeV, while it is consistent with the results of  Ref. \cite{ExpK0}.


%
\begin{table*}[t!]
\begin{tabular}{|c|c|c|c|c|c|c|c|c|c|c|c|c|c|c|c|}
                                                                                                      \hline
                                                       & \multicolumn{2}{|c|}{$\kappa=0.1$}
                                                       &\multicolumn{2}{|c|}{$\kappa=0.15$}&\multicolumn{2}{|c|}{$\kappa=0.2$}
                                                       &\multicolumn{2}{|c|}{$\kappa=0.25$}&\multicolumn{2}{|c|}{$\kappa=0.3$}      \\ \hline
                    $R~[fm]$                     &  0.28 &  0.42  & 0.35  &  0.48  & 0.41  &  0.50 & 0.47   &  0.52  & 0.53 & 0.54    \\ \hline
 $C_d^2~[MeV\cdot fm^{3\kappa}]$ &284.98&325.06&206.05&229.57&168.15&179.67&146.97&152.00&133.79&134.60 \\ \hline
                 $U_0~[MeV]$                  &567.32&501.65&343.93&312.83&231.42&217.76&162.03&157.41&114.32&113.84 \\ \hline
                 $K_0~[MeV]$                  &306.09&465.13&272.55&405.97&242.56&322.80&217.16&256.44&192.35&199.27 \\ \hline
                 $\mu_c~[MeV]$                &890.94 &881.01&900.08&895.08&906.44&904.49&911.11&910.53&914.74&914.70 \\ \hline 
                 $T_c~[MeV]$                  & 17.62 & 20.60 & 15.60 & 17.97 & 13.93& 15.36 & 12.49 &13.20 & 11.16 & 11.30  \\ \hline 
                             $n_c~[fm^{-3}]$               & 0.009  & 0.010 & 0.013 & 0.014& 0.016 & 0.017 & 0.018 & 0.020 &0.022& 0.022   \\ \hline 
       $p_c~[MeV\cdot fm^{-3}]$       & 0.0186 & 0.028 & 0.031 & 0.045& 0.043 & 0.055 & 0.053 & 0.061 &0.060& 0.062   \\   \hline 
       $Z_c~$       & 0.1173 & 0.1359 & 0.1529 & 0.1789 & 0.1929 & 0.2106 & 0.2357 & 0.2311 &0.2444 & 0.2494   \\   \hline 

\end{tabular}
\label{table1}
\caption{Different sets of parameters which simultaneously reproduce the properties of normal nuclear matter ($p=0$ and $n=n_0 = 0.16~fm^{-3}$
 at $\mu=923~MeV$, see text for details) and obey the proton  flow constraint on the nuclear matter EoS along with incompressibility factor $K_0$ 
and parameters of CEP. $R, C_d^2,  U_0$ and $\kappa$ are the adjustable  parameters of EoS, while the baryonic chemical potential $\mu_c$,  $T_c$, particle number density $n_c$, pressure $p_c$ and compressibility constant $Z_c \equiv \frac{p_c}{T_c\, n_c}$ at CEP 
are found  for each set of  model parameters.}
\end{table*}

The determined range of  $K_0$ and $T_c$ values allows us to reveal the mutual consistency of  experimental  results. 
Thus, the recent experimental estimates of the nuclear matter critical temperature  belong to the following range $15.5$ MeV $ \lesssim T_c \lesssim 21$ MeV \cite{Karnauch06,ExpT1,ExpT2,ExpT3}.  From Fig.  \ref{fig3} one can see that the values $T_c > 18$ MeV
are inconsistent with the upper range of  $K_0$ values, i.e.  the critical temperature values above $18$ MeV  require $K_0$ values above 315 MeV.   On the other hand the region   $15.5$ MeV $ \lesssim T_c \lesssim 18$ MeV
is consistent with the following range of values of incompressibility constant $K_0 \in [270; 315]$ MeV.  It is interesting that 
these  ranges of $T_c$ and $K_0$ values are consistent  with the inequality on the nucleon hard-core radius $R \le 0.35$ fm.
The latter  is just about 17\% above the  value   $r\simeq 0.3$ fm used in 
the realistic  nucleon-nucleon  interaction potential to reproduce   the low energy  nucleon-nucleon scattering data  \cite{Bohr,Andronic}.

Although the values of  $T_c$ and $K_0$  are very well consistent with the ones found for the RMF models  \cite{Delfino16,Menezes17},  the other characteristics of CEP, namely the pressure $p_c$, the particle number density $n_c$ and the compressibility  constant
$Z_c = \frac{p_c}{T_c\, n_c}$,  are essentially lower than the ones found by the RMF  as one can see from Table I.
Surprisingly,   the found $Z_c \in[0.117; 0.249]$ values demonstrate a rich diversity, but all of them   are in the range of values known for real liquids, namely $Z_c \simeq 0.117$ corresponds to 
the hydrogen fluoride, whereas $Z_c \simeq 0.249$ corresponds to 
the hydrogen chloride \cite{ChemicalConsts}. Among other real liquids which fall into the found range of  $Z_c$ values
we would mark the deuterium oxide ($Z_c \simeq 0.228$), ammonia ($Z_c \simeq 0.244$), water ($Z_c \simeq 0.229$), acetic acid ($Z_c \simeq 0.201$),
acetone ($Z_c \simeq 0.232$), acetonitrile ($Z_c \simeq 0.185$), metanol ($Z_c \simeq 0.223$) \cite{ChemicalConsts} etc. 
At the same time the range of   the critical compressibility constant  of  the  RMF models  is 
$Z_c^{RMF} \in [0.284; 0.331]$ \cite{Menezes17}, i.e. it is close or slightly above  the critical compressibility constants of the following substances   \cite{ChemicalConsts}
Ar, Kr, Xe, CH$_4$, N$_2$, O$_2$, and CO, but  there is no reason to believe that there is a close similarity  between the properties of  particularly   these atomic/molecular  gases and the gas  of nucleons. Therefore,  a priori for the realistic EoS one would expect an essentially wider spectrum of  $Z_c$ values, like the RMF  models show for $T_c$, $n_c$ and $p_c$ values. 

Of course, one may be  surprised by the low values of the critical  density found within the IST EoS, but we would like to remind the reader that all `experimental'  estimates of $n_c$ and $p_c$ are the model dependent ones.   Furthermore,  one should remember that 
our estimates for $n_c$ and $p_c$ correspond to a nuclear matter, while in the experiments one cannot ignore the Coulomb interaction.  Since there is no exact way to account for the  Coulomb interaction, then  an extraction of the  nuclear matter critical properties  is inevitably   model dependent procedure. 
{Moreover, it is  clear that, if in addition we  include into a model EoS with a fixed value  of $\kappa$ a repulsive Coulomb-like (i.e. weak)  interaction of large, but finite range, this would increase the  attraction strength $C^2_d$ to compensate the shift of binding energy.  This is apparent, since the long range repulsion will  affect the low density characteristics, namely  it will increase the pressure and binding energy per nucleon.
The increase of $C^2_d$   will, in turn,   increase the critical density and critical pressure (see the columns of same $\kappa$ values in Table I). Such a modification, however,  will  make the whole treatment too complicated and will  destroy the main attractive feature of this model, namely its simplicity. 
}

Besides,  the typical values of $n_c^{RMF}$  obtained in the RMF models  analyzed in \cite{Menezes17} are as follows  $n_c^{RMF} \in [0.295 \,n_0; 0.343\, n_0]$. Suppose that these are, indeed, the true values of  nuclear matter critical density. Then, if one included into  these EoS a repulsive Coulomb-like interaction of large, but finite range,  it immediately would  increase the critical density further, 
i.e. one would expect that the critical density in the real systems studied in experiments  should be larger than $n_c^{RMF}$. In this case, however, one faces 
a severe problem to explain how it comes that  the experimental data on size (charge) distribution of nuclear fragments demonstrate a power law which is typical for the CEP \cite{Karnauch06, Elliott,Reuter} and, moreover, how it comes that the statistical multifragmentation model  \cite{SMM} which up  to now is the most successful one in explaining the data obtained in the multifragmentation reactions  is able to reproduce the mentioned  power law with the break-up density  $n_{br} \simeq \frac{1}{6} n_0 - \frac{1}{3} n_0$ \cite{Karnauch06,SMM}? 
On the other hand, the low values of critical density obtained within the IST EoS do not face such a problem. Therefore, it seems that the  typical values of $n_c^{RMF}$ reported in  \cite{Menezes17} may evidence about some internal inconsistency of  these models.



\section{Conclusions} 
\label{Concl}

In this work we developed a novel family of EoS for symmetric nuclear matter based on the IST concept for the hard-core repulsion. It seems
that the quantum version of the IST EoS employed here catches the right physics, since having only four adjustable parameters
each formulation of this EoS is able to reproduce not only the main properties of the nuclear matter ground state (3 conditions), but, simultaneously,  it is able to obey  the proton flow constraint  \cite{Danielewicz} up to particle number density $0.75$ fm$^{-3}$ (at least 8 conditions).  Moreover, one  can easily check that all versions of the IST EoS presented here automatically obey the kaon production constraint  \cite{KaonConstr}.

A detailed analysis  of the proton flow constraint allows us to obtain the band of values for the incompressibility constant of normal nuclear matter $K_0$ and the critical temperatures $T_c$ which are consisted with the proton flow constraint. Assuming that the quantum IST EoS is valid up to the maximal packing fraction $\eta_{max} =0.2 $ and requiring that it holds for maximal particle number density of the proton flow constraint  $0.75$ fm$^{-3}$, we obtained the condition $R\le 0.4$ fm for the hard-core radius of nucleons. This condition rules out  the $K_0$ values below 250 MeV. 
Furthermore, analyzing the recent data on the critical temperature value $T_c \simeq 15.5-21$ MeV which, apparently, are not very accurate, we conclude that only the range $T_c \simeq 15.5-18$ MeV is consistent with the values  $K_0 \simeq 270-315$ MeV,
while the larger values of $T_c$ require $K_0$ values above $315$ MeV, which are not supported by the recent findings \cite{ExpK0}. 
It is interesting that the mutually consistent values of  $K_0$ and $T_c$ are 
also consistent with the inequality $R\le 0.35$ fm for the hard-core radius of nucleons. This is a remarkable finding since the value $0.35$ fm is just 17\% above the radius of nucleon-nucleon interaction potential and 
at the same  time this is just the hard-core radius of baryons found  recently by the IST formulation of the hadron resonance gas model from fitting the experimental hadron multiplicities measured in central nuclear collisions in the whole range of collision energies from $\sqrt{s_{NN}}=2.7$ GeV to $\sqrt{s_{NN}}= 2.76$ TeV \cite{Bugaev16,Sagun17,Bugaev17b}. Therefore, we conclude that the physically most justified range of these quantities is as follows: $K_0 \simeq 270-315$ MeV and $T_c \simeq 15.5-18$ MeV.  Based on these results, we hope that our systematic analysis of the correlations between the $K_0$ and $T_c$ values
will help to  establish  the mutual consistency of  their  values found with higher accuracy. 

The obtained hard-core radii of nucleons are essentially smaller than the ones found recently within the 
novel approach of Ref. \cite{Vovch17}. It seem that $R \ge 0.53$ fm claimed in  \cite{Vovch17}  are highly unrealistic, since in the IST EoS they correspond to very low values of $T_c \simeq 11.16-11.3$ MeV and $K_0 \simeq 192-199$ MeV (see the column $\kappa=0.3$ in Table I). It seems that such values   are generated by the   parameterization of internuclear attraction which is typical for ordinary liquids used  in  \cite{Vovch17}. Such a conclusion  is supported by a success of the mean-field  parameterization (\ref{IX}) employed here.
We would like to point out that the interaction pressure (\ref{IX}), as it was first found  in an old paper \cite{Gorenstein93}, cannot 
be expanded into a Taylor series at $n=0$ and, hence, the traditional virial expansion cannot be established for  this family of  IST EoS.  We hope that further studies of the EoS of  dense quantum liquids with strong interaction will clarify the question whether 
the non-analytic density dependence of pressure (\ref{IX}) is an inherent property of nuclear Fermi liquid or it is  common
for other Fermi liquids. 

In contrast to the RMF models, the developed family of EoS demonstrates a wide diversity of  values of the critical compressibility constant $Z_c$, namely $Z_c \simeq 0.117-0.249$, which, however, are well known for the ordinary liquids. Therefore, we hope it can be straightforwardly applied to
the quantum and classical liquids, to which the RMF models discussed here, apparently, cannot be applied.  \\

\noindent
{\bf Acknowledgments.}
The authors appreciate the valuable  discussions with V. Yu. Denisov and  I. N. Mishustin.
K.A.B., A.I.I. and V.V.S.  acknowledge  a partial  support from  
the Program of Fundamental Research of the Department of Physics and Astronomy of National Academy of Sciences of Ukraine.
 The work of K.A.B. and L.V.B. was performed in the framework of COST Action CA15213 ``Theory of hot matter and relativistic heavy-ion collisions" (THOR). K.A.B. is thankful to the  COST Action CA15213  for a partial support. 
 V.S. thanks the Funda\c c\~ao para a Ci\^encia e Tecnologia (FCT), Portugal, for the
partial financial support to the Multidisciplinary Center for Astrophysics (CENTRA),
Instituto Superior T\'ecnico, Universidade de Lisboa, through the Grant No.
UID/FIS/00099/2013. 
The work of L.V.B. and E.E.Z. was supported by the Norwegian Research Council
(NFR) under grant No. 255253/F50 - CERN Heavy Ion Theory.
 K.A.B. and  A.I.I. acknowledge a warm hospitality of the University of Oslo where this work was done.


\begin{thebibliography}{99}

\bibitem{Horst86}
H. St\"ocker  and W. Greiner, Phys. Rep. {\bf 137},  227 (1986).

\bibitem{Karnauch06}
V. A. Karnaukhov, Phys. Part. Nucl. {\bf 37}, 165 
(2006) and references therein.

\bibitem{Lattimer12}
 J. M.  Lattimer,  
 Annu. Rev. Nucl. Part. Sci.  {\bf  62}, 485 (2012)  and references therein.

\bibitem{StellarSMM13}
 N. Buyukcizmeci, A. S. Botvina, I. N. Mishustin,  
  Astrophys.\ J.\  {\bf 789}, 33 (2014).
  
\bibitem{Dutra14} 
M. Dutra et al., 
 Phys. Rev. C {\bf 90},   055203  (2014) and references therein.
 
\bibitem{Maslov17} 
E. E. Kolomeitsev, K. A. Maslov and D. N. Voskresensky,  
Nucl. Phys. A {\bf 961},  106 (2017).
 
 \bibitem{KAB07}
 %
 K. A. Bugaev, 
Phys. Part. Nucl.  {\bf 38}, 447 (2007). 
 
 \bibitem{BISO:13}
%
K. A. Bugaev, A. I. Ivanytskyi, V. V. Sagun and D. R. Oliinychenko, 
Phys.  Part. Nucl. Lett. {\bf 10}, 832 (2013).
  
\bibitem{Sagun2014a}
%
V. Sagun, A. Ivanytskyi, K. Bugaev, I. Mishustin, Nucl. Phys. A {\bf 924}, 24  (2014).     
 
\bibitem{Bugaev07rev}
 %
 K. A. Bugaev and P. T. Reuter,
 Ukr. J. Phys. {\bf 52},  489 (2007). 

 
\bibitem{Gulminelli15}
 %
S.Mallik, F. Gulminelli G. Chaudhuri, Phys. Rev. C  {\bf 92}, 064605 (2015).
 
  \bibitem{Delfino16}
 %
 O. Lourenco, B. M. Santos,  M. Dutra and A. Delfino,
  Phys.\ Rev.\ C {\bf 94},  045207 (2016).
 
 \bibitem{Menezes17}
 %
 O. Lourenco,  M. Dutra and D. P. Menezes, 
  Phys.\ Rev.\ C {\bf 95},  065212 (2017).

 
\bibitem{Danielewicz}
P. Danielewicz, R. Lacey and W. G. Lynch,  Science {\bf 298}, 1593 (2002). 
 
 
 \bibitem{Vovch17}
 %
 V. Vovchenko, 
  Phys.\ Rev.\ C {\bf 96},  015206 (2017)
  
\bibitem{Bugaev2017}
K. A. Bugaev et al., arXiv: 1704.06846 [nucl-th].  

\bibitem{ExpK0}
 %
 J. R. Stone, N. J. Stone and S. A. Moszkowski, Phys.
Rev. C {\bf 89}, 044316 (2014). 

\bibitem{Gorenstein93}
M. I. Gorenstein et al., J. Phys. G {\bf 19}, 69 (1993).

\bibitem{Rischke91}
D. H.  Rischke,  M. I.  Gorenstein,  H.  St\"ocker and  W.  Greiner, Z. Phys. C {\bf 51}, 485 (1991).

\bibitem{Bugaev16}
K. A. Bugaev et al.,  arXiv:1611.07349v2 [nucl-th].

\bibitem{Sagun17}
V. V. Sagun et  al.,  arXiv:1703.00049 [nucl-th].

\bibitem{Bugaev17b}
K. A. Bugaev et al.,  arXiv:1709.05419 [hep-ph].


\bibitem{Zeeb08}
G. Zeeb, K. A. Bugaev, P. T. Reuter and H. St\"ocker,
Ukr.  J. Phys. {\bf 53},   279 (2008).

\bibitem{KaonConstr}
W. G. Lynch et al., 
Prog. Part. Nucl. Phys.  {\bf 62}, 427 (2009).


\bibitem{ExpT1}
 %
V. A. Karnaukhov et al., Nucl. Phys. A {\bf 780}, 91 (2006). 

\bibitem{ExpT2}
 %
V. A. Karnaukhov, Phys. At. Nucl.  {\bf 71}, 2067 (2008). 

\bibitem{ExpT3}
 %
J. B. Elliott, P. T. Lake, L. G. Moretto and L. Phair,
Phys. Rev. C {\bf 87}, 054622 (2013).

\bibitem{Bohr}
see, e.g., A. Bohr and B. Mottelson, ``Nuclear  Structure" (Benjamin, New York 1969), Vol.1.

\bibitem{Andronic}
 A. Andronic, P. Braun-Munzinger and J. Stachel, Nucl. Phys. A {\bf 772}, 167 (2006) and references therein.

\bibitem{ChemicalConsts}
 %
Kaye\&Laby, Tables of Physical and Chemical Constants, Sect.3.5, National Physical Laboratory,  UK, 
 http://www.kayelaby.npl.co.uk/chemistry/3\_5/3\_5.html

\bibitem{SMM}
 %
J. P. Bondorf et al., Phys. Rep.  {\bf 257}, 131 (1995) and references therein.

\bibitem{Elliott}
 %
J. B. Elliott et al., EOS Collaboration, Phys. Rev. C {\bf 62}, 064603 (2000).

\bibitem{Reuter}
 %
P. T. Reuter and K. A. Bugaev, Phys. Lett.  B {\bf 517},  233  (2001).


\end{thebibliography}
\end{document}